\documentclass[sn-nature]{sn-jnl}% Style for submissions to Nature Portfolio journals
\usepackage{graphicx}%
\usepackage{multirow}%
\usepackage{amsmath,amssymb,amsfonts}%
\usepackage{amsthm}%
\usepackage{mathrsfs}%
\usepackage[title]{appendix}%
\usepackage{xcolor}%
\usepackage{textcomp}%
\usepackage{manyfoot}%
\usepackage{booktabs}%
\usepackage{algorithm}%
\usepackage{algorithmicx}%
\usepackage{algpseudocode}%
\usepackage{listings}%
\usepackage{siunitx}
\usepackage{comment}
\usepackage{lineno}
\usepackage{subfig}
\usepackage{todonotes}
\usepackage{lmodern} %Fixes the 'Font shape XXX not available' error

\begin{document}

\title[On the application of components manufactured with stereolithographic 3D printing in high vacuum systems]{On the application of components manufactured with stereolithographic 3D printing in high vacuum systems} %Title of paper


%\preprint{}

\author*[1]{\fnm{Aleksandar} \sur{Radi\'{c}}}\email{ar2071@cam.ac.uk}

\author[1,2]{\fnm{Sam Morgan} \sur{Lambrick}}\email{sml59@cam.ac.uk}

\author[2]{\fnm{Sam} \sur{Rhodes}}\email{s.rhodes@ionoptika.co.uk}

\author[1,2]{\fnm{David James} \sur{Ward}}\email{djw77@cam.ac.uk}

\affil[1]{\orgdiv{Department of Physics}, \orgname{University of Cambridge}, \orgaddress{\street{19 J.J. Thomson Avenue}, \city{Cambridge}, \postcode{CB3 0HE}, \country{United Kingdom}}}
\affil[2]{\orgname{Ionoptika Ltd.}, \orgaddress{\street{Units B5-B6, Millbrook Close}, \city{Chandlers Ford}, \postcode{S053 4BZ}, \country{United Kingdom}}}


\abstract{We explore the ultrahigh-vacuum (UHV) compatibility of Formlabs `Clear Resin' \textit{via} vat photopolymerization (VPP). We report on a method for using VPP additive manufacturing, specifically Formlabs' widely available stereolithographic (SLA) printing using their `Clear Resin' material, to rapidly and cheaply prototype components for use in high-vacuum (HV) environments. We present pump down curves and residual gas analysis to demonstrate the primary vacuum contaminant from freshly printed SLA plastics is water with no evidence of polymers outgassing from the material and thus the vacuum performance can be controlled with simple treatments which do not involve surface sealing. An unbaked vacuum system containing SLA printed components achieved \SI{1.9e-8}{\milli\bar} base pressure whilst retaining structural integrity and manufacturing accuracy. Outgassing rates in the HV test chamber and preliminary results in a UHV chamber indicate that our method can be extended to achieve ultrahigh-vacuum compatibility. We further report on the effect of atmospheric exposure to components and present evidence to suggest that water re-ad/absorption occurs exclusively on the surface, by showing that the bulk mass changes of the material is irreversible on the timescale investigated ($< 2\,$weeks).}
%\keywords{keyword1, Keyword2, Keyword3, Keyword4}
%\pacs{}% insert suggested PACS numbers in braces on next line

%\maketitle %\maketitle must follow title, authors, abstract and \pacs


\maketitle
\section{Introduction}
\label{sec1}
Additive manufacturing (AM) is a rapid fabrication technique with a large amateur and industrial user base. Printer sales are growing at more than 20\% per annum \cite{3DPrintingMarketSize} and projected sales of 21.5 million units in 2030. Additive manufacturing benefits from wide applicability in small-scale manufacturing, prototyping and refining of new components due to fast turn around times, capability to fabricate internal and external structures impossible to fabricate using conventional techniques, low per-unit costs, efficient tool-chain from concept to component and wide material choice. These benefits are particular relevant to research \& development and small volume-high value manufacturing, two areas where the use of (ultra)high-vacuum equipment is ubiquitous. Several printing technologies are available in the marketplace from large industrial machines to desktop devices, in addition different print media can be used to suit precision and material properties required.

The application of AM components in high-vacuum systems has been limited. There are reports discussing the use of both green and cured components manufactured using material extrusion (ME) and vat photopolymerization (VPP). ME components have been shown to have potential for (ultra)high-vacuum compatibility depending on the ultimate surface finish, porosity and specific material used, achieving pressures $>\SI{1e-7}{\milli\bar}$, with typical outgassing rates between $\SI{e-5}{}-\SI{e-7}{\milli\bar\litre\per\second\per\centi\meter\squared}$ reported for VPP-based polymers\cite{Gans2014,Povilus2014,Zwicker2015,Mayville2022,Sun2017,Peacock1980}. Other VPP-based materials, such as ceramics, have also been shown to achieve high-vacuum compatibility, reaching $\sim\SI{2e-7}{\milli\bar}$ base pressures\cite{Yang2018,Eckhoff2023,IzquierdoReyes2022}. Previous work has also explored the application of vacuum sealants, such as ``VacSeal'' from SPI Supplies Inc.\cite{StructureProbeInc.}, post-manufacturing on ME-based acrylonitrile butadiene (ABS) and polycarbonate (PC) components , to create a physical barrier through which trapped water cannot out-gas\cite{Chaneliere2017,HEIKKINEN2020125459,jmmp6050098}. One could increase the efficacy of coating methods by smoothing the material's surface, for example ABS can be smoothed using an acetone dipping bath or vapor exposure. Designed primarily as leak sealants for high- and ultrahigh-vacuum systems, such products typically contain toxic solvents that affect repeatability and present environmental challenges. Crucially, application of a surface coating limits the degree of internal complexity of printed components as any surface left un-coated will result in performance limited outgassing. Many common vacuum sealants require high temperature curing, 260 - 300 \textdegree C for VacSeal specifically\cite{StructureProbeInc.}, which are not compatible with many materials used for fabrication. Typical pressures achieved using leak sealant coatings range between $\num{e-4} - \num{e-7}\,$mbar when applied to ABS or PC\cite{Chaneliere2017,HEIKKINEN2020125459,jmmp6050098,7466442}.

VPP-based methods have become a standard method due to the control and precision of manufacturing and is substantially different to it's stablemates since the finished component is closer to a chemically bonded solid body and structures are not defined by lamination layers, thus VPP is a promising candidate for producing vacuum compatible bespoke components. VPP printers, and their resins, are available at low cost and are readily accessible to the public, both to own operate personally, and \textit{via} third party services. Layers of $10-\SI{20}{\micro\metre}$ with overall component tolerances of $\pm \SI{0.15}{\micro\metre}$ are commonplace in the consumer market (e.g. 3Dhubs.com), and contract printing services are readily accessible at low unit  cost for one-off prints or small manufacturing runs.

In the current work, we investigate the vacuum properties of Formlabs `Clear Resin' polymer manufactured \textit{via} VPP, henceforth referred to using the commercial name `stereolithographic (SLA)' printing\cite{clear_data}, and show that effect of the bulk water can be mitigated without vacuum sealants, reaching the test vacuum chamber base pressure of \SI{1.9e-8}{\milli\bar}. We define a post-manufacture method to process SLA plastics to make them compatible in high- and ultrahigh-vacuum systems (HV and UHV, respectively), alongside characterisation of their water re-ad/absorption properties upon exposure to atmosphere. Following the outlined procedure it is possible to use SLA printed components in HV systems whilst retaining structural integrity and manufacturing tolerances. While there have been few reported usages of SLA components in HV systems, recent usage, such as that of the complex optical components for neutral helium microscopy \cite{Radic3D,Bergin_2021,LambrickDiffuse,von_jeinsen_2d_2023,defects_radic,2d_methods}, quadrupole mass spectrometers \cite{Eckhoff2023,IzquierdoReyes2022} and vacuum pump components\cite{Taylor2017}, indicate that there is a wider interest in the application of SLA components in HV systems beyond vacuum science itself.

%\section{Sample Information}
To investigate the outgassing and water re-ad/absorption properties of SLA 3D printed plastic, a regular tetrahedron (nominal $\SI{4.80}{\centi\metre}$ edge length, $\sim\SI{13}{\centi\metre^3}$ volume) was chosen as the test sample. Images of the test samples are shown in figure \ref{fig:sample_images}. A tetrahedron was chosen to maximize the ratio of volume to surface area to establish a worst-case scenario where a vacuum component has the largest possible internal volume of water, and is of typical size for a vacuum component one would want to produce using additive manufacturing. Six samples were printed by 3DHubs \cite{3dhubs} using Formlabs ``Clear Resin'' \cite{plastic1,clear_data} with a Formlabs ``Form 3'' printer costing approximately £\,5 per sample with a lead time of 3 days. While the samples were ordered from 3DHubs, the chosen resin and printer are standard and representative of the most widely available SLA additive manufacturing services and methods. We also expect that the findings presented are broadly applicable to other VPP-based polymers. Upon delivery the samples were within the stated dimensional tolerance ($\pm \SI{0.15}{\milli\metre}$) with smooth, flat faces, and sharp edges with the exception of small aberrations on the side where a support structure attached during fabrication.

\section{Method}
The samples  were baked in a ultrahigh-vacuum test chamber with UHV base pressure ($\sim\SI{e-10}{\milli\bar}$). Vacuum properties were investigated in a separate vacuum system using a Hiden Analytical residual gas mass spectrometer (HAL/3F RC301 PIC300) \cite{HidenAnalyticalSpecs}. Schematic diagrams of both vacuum systems can be found in Figures \ref{fig:hv_chamber_schematic}, \ref{fig:uhv_chamber_schematic}. The atmospheric exposure time was varied to explore the vacuum properties of the sample with respect to water re-ad/absorption from atmosphere. Atmospheric temperature and relative humidity were recorded using a ``PICO Humidiprobe''\cite{humidiprobeData}. The post-manufacture sample preparation, baking and vacuum property measurement procedures are outlined below.

\begin{enumerate}
    \item Clean the surface of the sample  by wiping with isopropyl alcohol and dry immediately using compressed dry air.
    \item Record mass and dimensions of samples.
    \item Bake samples in a vacuum oven (referred to as the ultrahigh-vacuum test chamber in this work). Increment the temperature by \ang{20}C steps every half hour until \ang{120}C is reached, bake for 48 hours. Allow samples cool slowly by turning off heating element and allowing for natural cooling within the oven. Gradual heating/cooling prevents structural distortion from material and water expansion.
    \item On removal from oven, visually inspect the samples and measure dimensions. Transfer samples to either the test vacuum chamber or store in an opaque storage container with temperature and humidity monitoring\cite{humidiprobeData}.
    \item Planned atmospheric exposure times, in the storage container, of 0, 1, 2, 4 and 7 days such that each sample is exposed for a different length of time.
    \item Record pump down curves for the samples in the test vacuum chamber.
    \item Acquire mass spectra using residual gas analyser \cite{HidenAnalyticalSpecs} immediately before venting for next sample. Use mass spectra to find the pressure contribution of water outgassing from each sample by integrating the spectrum to the total pressure.
    \item Repeat steps 4-7 for all samples.
\end{enumerate}
The samples were all placed in the ultrahigh-vacuum test chamber, as shown in figure \ref{fig:sample_images}, and heated to $\SI{120}{\celsius}$ in $\SI{20}{\celsius}$ steps every half hour. Oven pressure was initially \SI{4e-5}{\milli\bar} at room temperature, compared to $<\SI{1e-9}{\milli\bar}$ without any sample, rising to above the accurate range of an Edwards wide range gauge (WRG) \cite{EdwardsGaugeSpecs}. Bake temperature was sustained for \SI{48}{\hour} after which the chamber gradually cooled over approximately \SI{3}{\hour}.

Once the oven was vented to air, the 0 day sample was weighed and transferred, using gloves and in atmosphere, to the test chamber in less than 10 minutes. The other samples were also weighed and placed in an opaque, temperature and humidity monitored container for storage. The samples were stored at $\SI{21.5}{\degreeCelsius}$ and $\SI{32.7}{\percent}$ relative humidity with standard deviations of $\SI{0.6}{\percent}$ and $\SI{1.1}{\percent}$, respectively, over the two week experimental period. The mean mass lost during baking was $0.29\pm0.005\%$ from the initial $17.27\pm0.05\,$g mean. Sample dimensions were measured using electronic calipers pre- and post-baking, finding that the side lengths of the tetrahedra remained $\SI{4.80}{\centi\meter}$, the same as the nominal dimension to within the $\pm\SI{0.15}{\milli\meter}$ manufacturing tolerance.

All analysis of vacuum properties were conducted in a test chamber (\SI{20}{\litre} volume with nominal \SI{450}{\litre\per\second} pumping capacity) with a Hiden Analytical mass spectrometer used for residual gas analysis connected to the sample area via a UHV gate valve such that the ion source for the mass spectrometer remained on for consistency of measurements. The test chamber is vented to air for each sample change. A schematic diagram of the test chamber vacuum system is shown in Figure \ref{app:hv_chamber_schematic}.

\section{Results}
\begin{figure}[h]
    \centering
    \includegraphics[width = 0.75\linewidth]{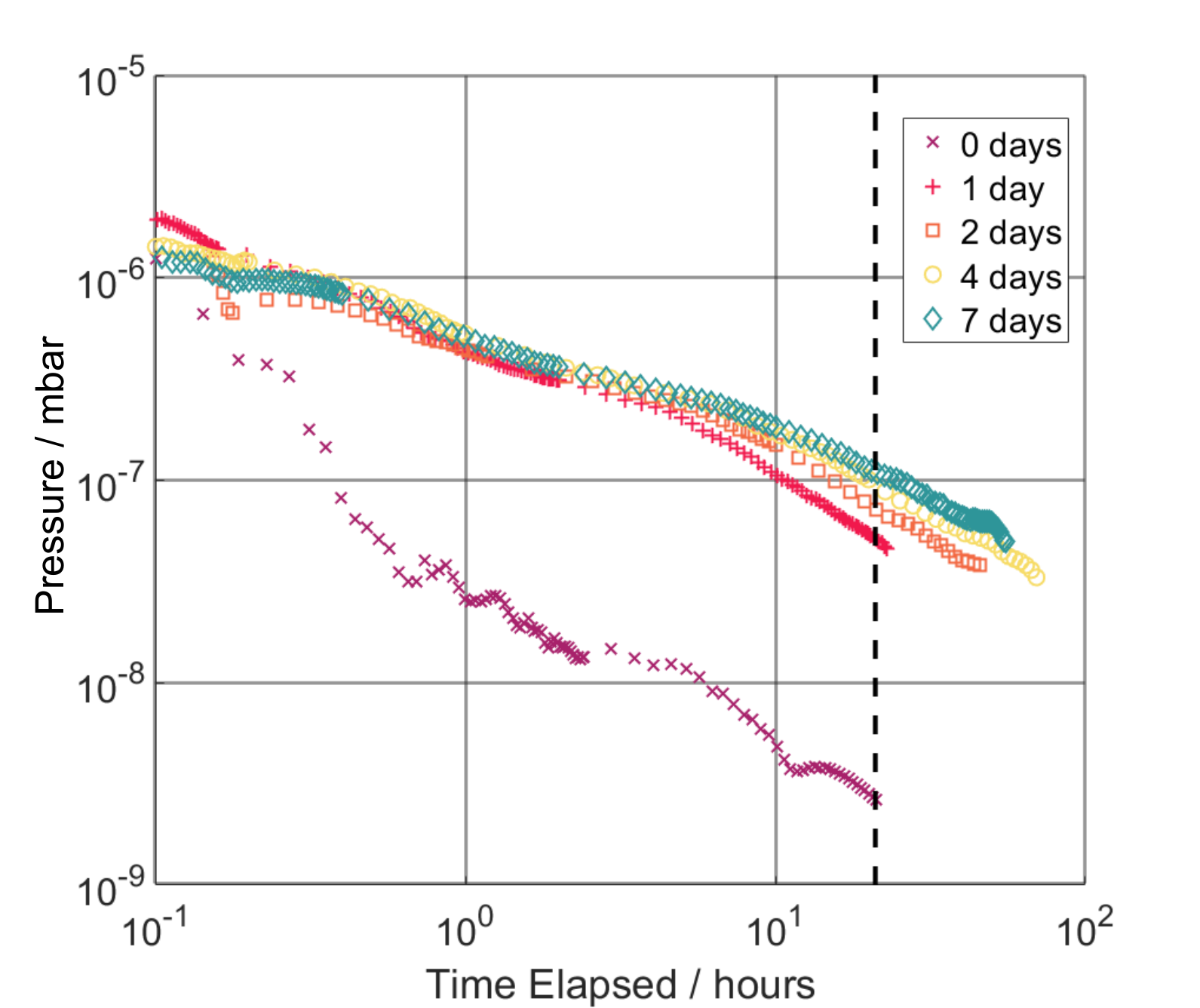}
    %,natwidth=1018,natheight=855
    \caption{Pressure contributions for samples exposed to atmosphere for between 0-7 days. An empty chamber pumping curve was subtracted from the samples' pumping curves to estimate pressure contribution of the samples to the vacuum. Vertical slice (dashed line) through the pumping curves is shown in Figure \ref{fig:outgassing_abs} to illustrate the dependence of base pressure on exposure time.}
    \label{fig:outgassing_plot}
\end{figure}

Pump down curves for each sample were recorded to demonstrate the water re-ad/absorption effect caused by atmospheric exposure over time. The pressure contribution, which can be equivalently expressed as outgassing per unit area, of each sample was calculated by subtracting the empty chamber pump down curve from that of the chamber with each sample. The pressure contribution for the samples exposed to air for different duration are shown in Figure \ref{fig:outgassing_plot}. Pressure contributions of the samples are alternatively provided in terms of outgassing per unit area in Table \ref{tab:outgassing_rates} along with a brief discussion on the major sources of uncertainty.

\begin{figure}[h]
    \centering
    \includegraphics[width = 0.75\linewidth]{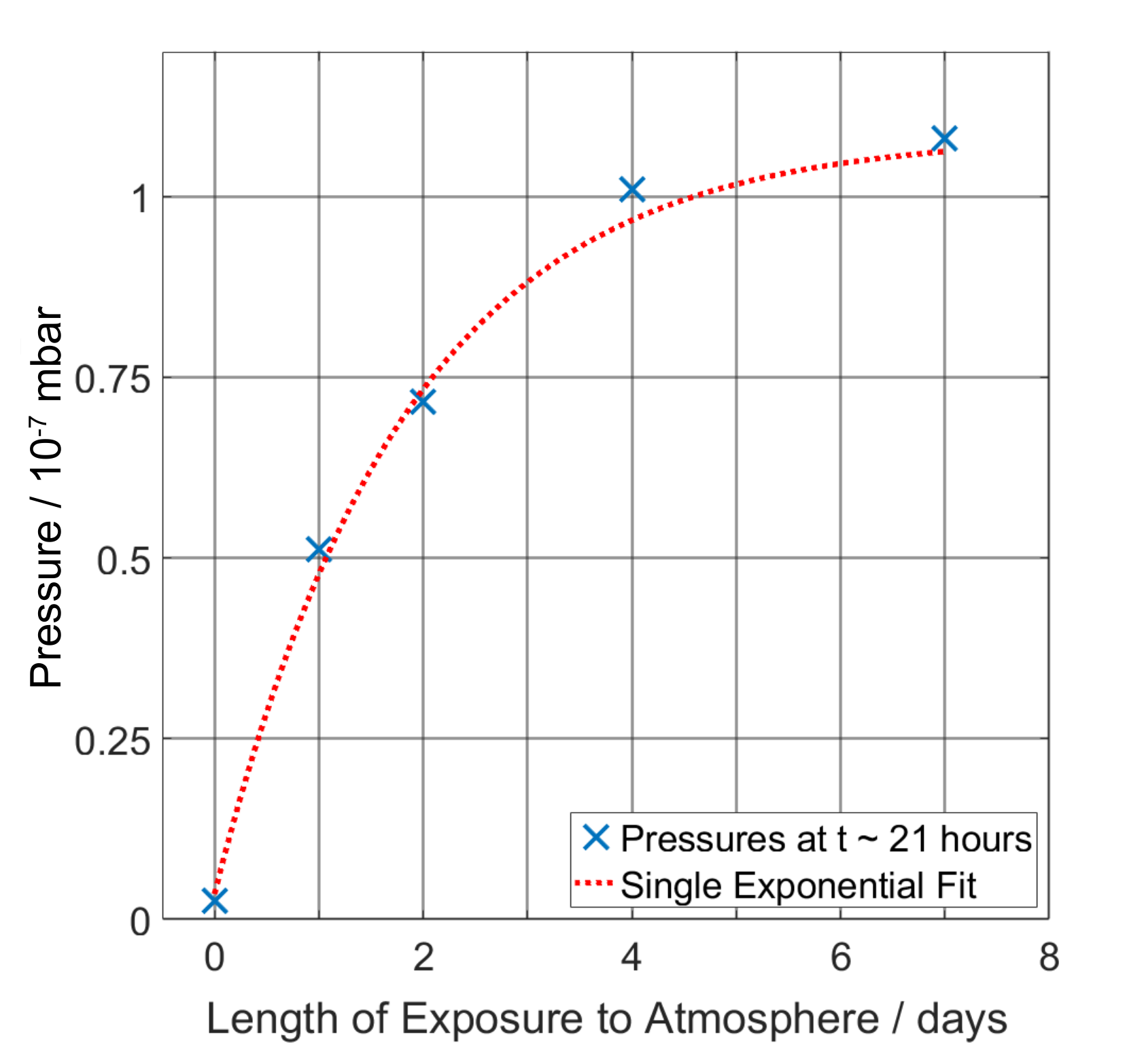}
    %, natwidth=1077,natheight=1002
    \caption{Pressure contributions as a function of exposure time taken from a vertical slice through Figure \ref{fig:outgassing_plot} at $t = \SI{21}{\hour}$. A single exponential with time constant $\tau = \SI{-0.70}{days}$ accurately describes the water re-ad/absorption rate under atmospheric conditions with $R^2 = 0.988$. Pressures are alternatively presented as specific outgassing rates in Table \ref{tab:outgassing_rates}.}
    \label{fig:outgassing_abs}
\end{figure}

Figure \ref{fig:outgassing_plot} plots the pressure in the test chamber due to each sample, calculated by subtraction of a reference pump down curve from that of a given sample. The figure shows a strong correlation between exposure time and water re-ad/absorption of the samples, with the majority of the water re-ad/absorption occurring in the first 24 hours of exposure. We assume that the majority of the re-ad/absorbed vacuum contaminants are water molecules, although we show that hydrogen is also present in the baked samples' pressure contribution. Taking a vertical slice through the figure at $\sim 21$ hours elapsed time yields the sample outgassing as a function of atmospheric exposure time, shown in Figure \ref{fig:outgassing_abs}.

Figure \ref{fig:outgassing_abs} shows that the water re-ad/absorption is very well described by a single exponential, showing that almost half the total water re-ad/absorption occurs within the first 24 hours of exposure, making a matter of hours the critical time period for exposure to achieve high-vacuum in a pump down time comparable to an empty chamber. 

A pressure of \SI{1.9e-8}{\milli\bar} was achieved after 21 hours pumping down with the 0 day sample in the high-vacuum (unbaked) test chamber, and \SI{9.9e-10}{\milli\bar} in the ultrahigh-vacuum test chamber using sample transfer in air. The pressures achieved in the respective test chambers demonstrate that Formlabs `Clear Resin' printed using SLA is compatible with HV and UHV systems when the presented baking method is applied. 

\begin{figure}[h]
    \centering
    \includegraphics[width = 0.75\linewidth]{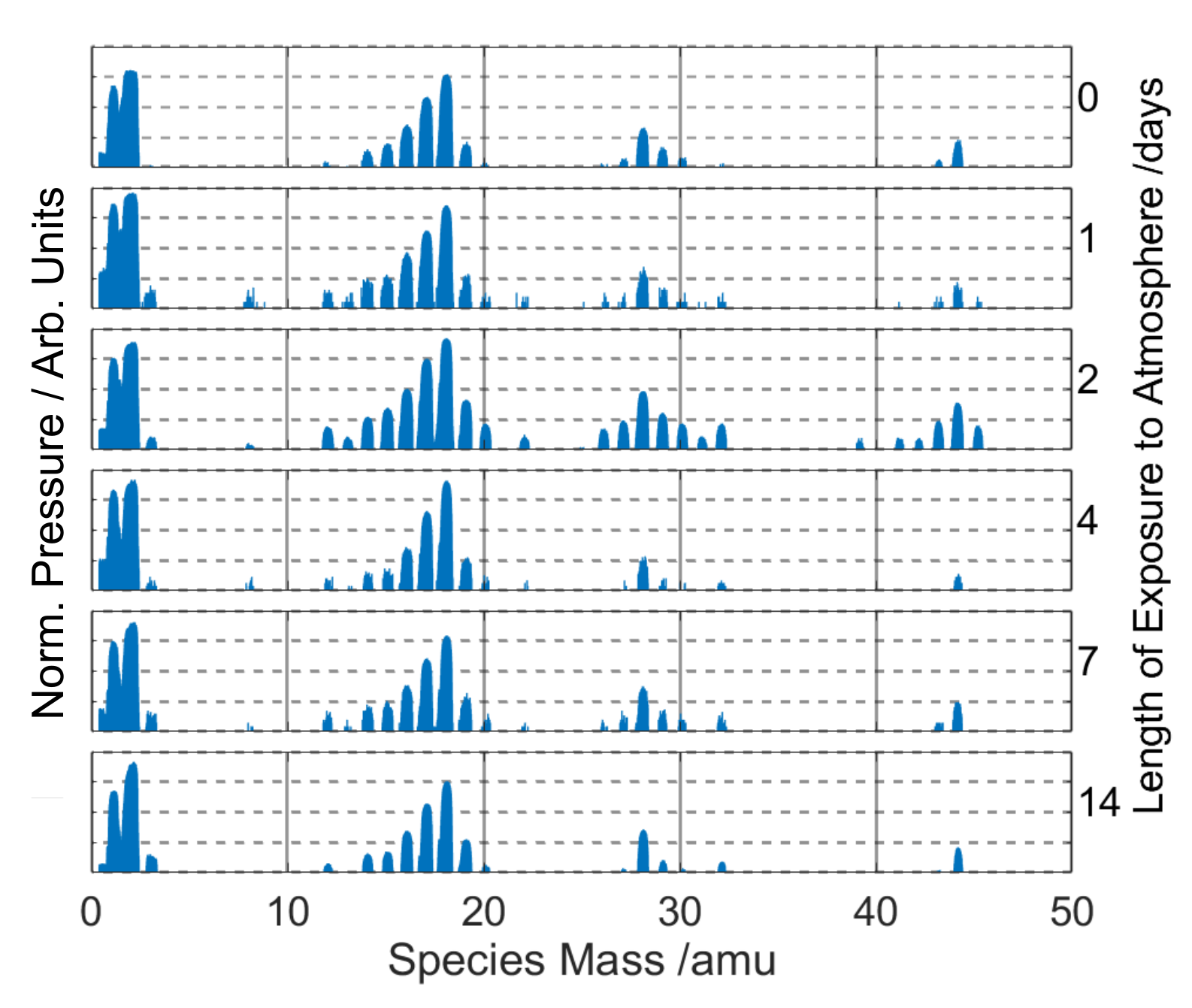}
    %h,natwidth=1015,natheight=858
    \caption{Mass spectra of samples exposed to atmosphere for 0-14 days (right side y-axis labels) were normalised to their respective base pressures achieved (Figure \ref{fig:outgassing_plot}) to show that the vacuum properties of all baked samples are consistent provided pump down times approximately equal to the time the samples were exposed to the atmosphere. All spectra have the same logarithmic y-axis range.}
    \label{fig:stacked_rga_tetra}
\end{figure}

Figure \ref{fig:stacked_rga_tetra} presents mass spectra taken for each sample after 21 hours, corresponding to data points in Figure \ref{fig:outgassing_abs} and the dashed line in Figure \ref{fig:outgassing_plot}. The mass spectrum of each sample was normalised to the respective pressure contributions after 21 hours pumping down to distinguish between chemical composition of vacuum contaminants after varying exposure to air. Mass spectra have been cropped at 50 amu because there was no significant detection of heavier species, indicating that the SLA plastic itself, and any cracking products, do not evolve into the vacuum.

The similarity in the distribution of species present in the normalised mass spectra in Figure \ref{fig:stacked_rga_tetra} indicates that, although the pump down speed slows as exposure time increases, water re-ad/absorption is limited to the surface on the time scales investigated. If the water re-ad/absorption of baked samples  was a bulk process, one would expect water to permeate deep into the sample such that it cannot be quickly removed by heating. Baking effectively removes water from the bulk of the material and prevents water from reabsorbing into it. Therefore, although the majority of water re-ad/absorption occurs in the first week of exposure, water can only saturate the surface which functions as a finite volume of water when outgassing into vacuum. In contrast to the unbaked plastic whose bulk water content acts as an infinite volume of water which cannot out-gas on a practical time-scale of days to weeks, hence the base pressure of only \SI{4e-7}{\milli\bar} achieved in the first instance after almost 2 weeks pumping down, in comparison to \SI{1.9e-8}{\milli\bar} in under $\SI{24}{\hour}$ for a baked sample. Note that the pump down time to achieve such pressures with baked samples  depends on time exposed to air, as discussed previously.

\begin{figure}[ht]
    \centering
    \includegraphics[width = 0.75\linewidth]{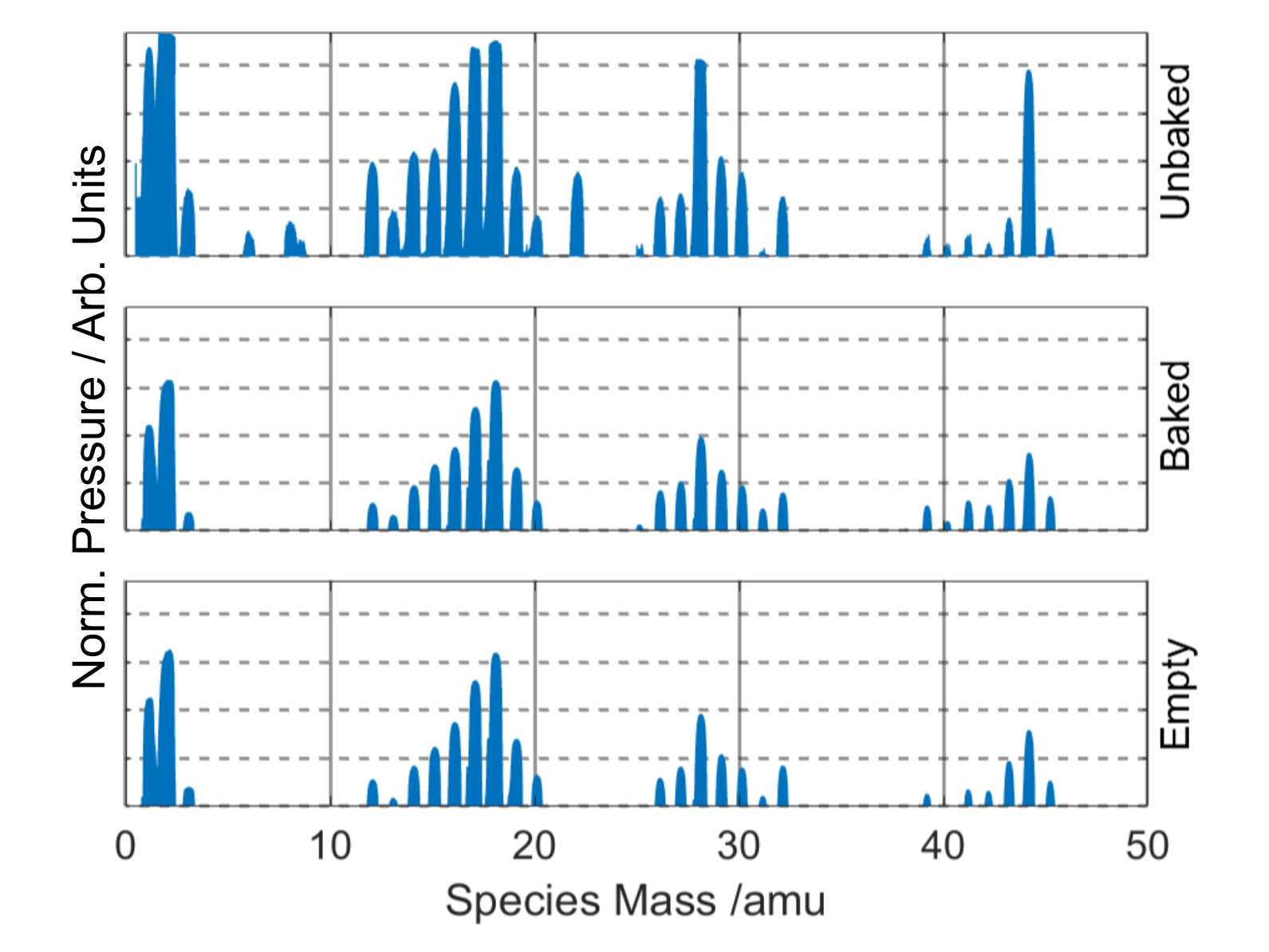}
    %, natwidth=1018,natheight=764
    \caption{Mass spectra, normalised as in Figure \ref{fig:stacked_rga_tetra}, for an unbaked sample (top), baked sample with 0 day exposure time (middle) and an empty chamber (bottom). All spectra have the same logarithmic y-axis range. The spectra for an empty chamber and baked sample are near identical, both at the base pressure of \SI{1.9e-8}{\milli\bar}. It should be noted that the electron multiplier saturated during the unbaked measurement, taken at \SI{4e-7}{\milli\bar}, providing lower bounds for the 2, 18 and 28 amu peaks.}
    \label{fig:stacked_rga_plates}
\end{figure}

Taking the chemical composition of Formlabs ``Clear Resin'' as $75\%$ poly(methyl methacrylate) (PMMA) and $25\%$ hydroxypropyl methacrylate (HPMA) by mass according to its safety data sheet \cite{clear_data}, the total water weight, referred to as weight in weight (w/w), pre-baking can be approximated. PMMA is reported to hold up to $2\%$ w/w water, with HPMA holding $0.2\%$\cite{pmmaWater,JamorinInternationalLtd.,OpesInternationalLtd.}. Averaging the water w/w values yields an overall max water content $\sim1.6\%$ for Formlabs ``Clear Resin''. If one takes the $0.29\%$ average mass loss during baking to be solely water based on mass spectra in Figure \ref{fig:stacked_rga_plates}, then $0.29/1.6\approx20\%$ of the maximum initial water mass was degassed due to baking. One can approximate the mass of water that is adsorbed to the surface of the sample by taking that generic plastics have roughly 200-400 monolayers of water adsorbed to their surface under atmospheric conditions. Taking the upper-bound value of 400 monolayers, we approximate the volume of surface water to be $\sim\SI{2e-4}{\per\centi\meter\cubed}$, giving the total surface water as $\SI{2e-4}{\gram}$, or $\SI{0.05}{\percent}$ of the mass lost during baking assuming complete desorption. Thus confirming that the majority of the water lost during baking is from the bulk of the sample. Importantly, this means that approximately $\SI{80}{\percent}$ of the total initial water mass remains in the bulk of the sample post-baking and is unable to diffuse through the plastic to provide an upper-bound outgassing rate $<\SI{2.9e-8}{\milli\bar\litre\per\second}$ in the HV chamber, and $<\SI{2.3e-9}{\milli\bar\litre\per\second}$ in the UHV chamber. Our results represent a significant improvement over reported outgassing rates for VPP-based polymers, typically $\SI{e-5}{}-\SI{e-7}{\milli\bar\litre\per\second\per\centi\meter\squared}$. A tabulated list of upper-bound outgassing rates for each sample as a function of exposure time is contained in table \ref{tab:outgassing_rates}, with the calculation method in Appendix \ref{app:outgassing_rates}.

Figure \ref{fig:stacked_rga_plates}  presents normalised mass spectra for an unbaked and baked sample, with 0 day exposure time, with the empty chamber below, 21h after evacuation. It is clear that water is the principle species introduced from the untreated plastic, and it can been seen that the vacuum properties of a baked sample are indistinguishable from the empty chamber to pressures lower than \SI{1.9e-8}{\milli\bar}.

A further experiment was conducted to evaluate the ultrahigh-vacuum compatibility of SLA Clear Resin by baking out the ultrahigh-vacuum test chamber with the sample inside to give a true UHV platform in the $\num{e-10}\,$mbar range such that the achievable pressure with no atmospheric exposure of the sample  could be established. The same baking procedure presented earlier was followed except the maximum temperature was increased to \ang{170}C. The pressure achieved with the sample  \textit{in situ} was indistinguishable to that without the sample  installed, validating the potential for using Formlabs Clear Resin in UHV systems where a moderate temperature bake may be used. However it must be noted that upon removal from the chamber, the surface of the sample developed shallow cracks which can only be attributed to the higher temperature used causing water vapour to escape destructively. After preliminary testing of the baking procedure at $100$, $120$, $\SI{170}{\degreeCelsius}$, we recommend gentle heating at a rate of $\SI{40}{\degreeCelsius\per\hour}$ and a maximum bake temperature of $\SI{120}{\degreeCelsius}$ be used for the presented material. $\SI{100}{\degreeCelsius}$ required noticeably longer baking to achieve similar performance, and $\SI{170}{\degreeCelsius}$ performed well in vacuum but caused the material to crack at the surface, pictured in figure \ref{fig:sample_images}. It is possible that the cracking could be avoided by heating more slowly than suggested here, or by using a different material, but this requires further investigation.

\section{Conclusion}
In the current work we have demonstrated that Formlabs Clear Resin polymer manufactured \textit{via} VPP is an appropriate material and manufacturing method combination for the cheap and rapid prototyping of small high-vacuum components with complex internal geometries, without application of surface sealants post-manufacturing. A simple baking protocol has been designed and tested that achieves (ultra)high-vacuum compatibility with VPP printed Formlabs Clear Resin, we anticipate that the result reflects the general performance of VPP-based polymers. We have shown that Formlabs Clear Resin is a favorable alternative to previously reported ceramics ,that are also VPP-based, when absolute pressure is required over specific mechanical or thermal properties\cite{Eckhoff2023,IzquierdoReyes2022}.

We demonstrate that the surface of SLA plastics re-adsorbs water quickly under room temperature and humidity atmospheric conditions. We recommend that SLA components be kept in a dehumidified or vacuum environment for short term storage post-baking to minimise vacuum chamber pump down times. Additionally, evidence is presented to suggest that water re-ad/absorption is exclusively a surface process, making the change in bulk water content due to baking permanent over the time scale investigated. Further investigation into $> 2\,$week atmospheric exposures is needed to explore the rate and extent to which the bulk of the plastic can reabsorb water to determine whether baking SLA plastics permanently changes their mass composition, and therefore induces permanent HV and UHV compatibility.

Some limitations have been found for higher temperature baking of SLA printed plastics at \ang{170}C which would reduce the treatment time and potentially further improve base pressure by extracting more water in a timely manner. Surface cracking of the plastic was observed  which we attribute to the water desorption occurring explosively, we can't conclusively show that the ultimate temperature, heating and cooling rates and post-manufacturing curing processes are the dominant contributor to that process, however there is clearly further capacity to investigate and improve all these aspects of the process for more demanding vacuum requirements. Additionally, high-temperature and high-strength variants of SLA plastics are also available which may be more resistant to cracking from rapid water degassing during baking. It is likely that dehydration of components will cease to be the dominant factor and that other out gassing products will be the ultimate limitation to ultimate vacuum performance. It is possible that manufacturing components in a low-humidity/vacuum environment may also improve ultimate vacuum properties. Taking the typical atmospheric humidity as $30\%$ during both printing and post-bake exposure, the water content of air is $\sim\SI{9500}{ppm}$. In comparison, a dry glovebox typically achieves water concentration $<\SI{0.1}{ppm}$. If this difference in ambient water concentration would be reflected in the composition of the final component, recalling that we calculate a $-20\%$ change in total water mass due to baking, the plastic may be immediately HV compatible and exhibit water re-ad/absorption properties similar to that of the baked components explored in this study.

\section*{CRediT authorship contribution statement}
\textbf{Aleksandar Radi\'{c}:} Writing - original draft, Methodology, Investigation, Data curation, Visualization, Formal analysis. \textbf{Sam Morgan Lambrick:} Writing - review \& editing, Supervision. \textbf{Sam Rhodes:} Writing - review \& editing, Project administration.  \textbf{David James Ward:} Supervision, Resources, Funding acquisition.

\section*{Declaration of competing interest}
Sam Rhodes reports a relationship with Ionoptika Ltd. that includes: employment. David Ward reports a relationship with Ionoptika Ltd. that includes: consulting or advisory. The other authors declare that they have no known competing financial interests or personal relationships that could have appeared to influence the work reported in this paper.

\section*{Data Availability}
All supporting data is available in the Supplementary Information and, upon publication, at the Apollo (University of Cambridge repository) \url{https://doi.org/10.17863/CAM.104367}.

\section*{Acknowledgements}
The work was supported by EPSRC grant EP/R008272/1, Innovate UK/Ionoptika Ltd. through Knowledge Transfer Partnership 10000925.The work was performed in part at CORDE, the Collaborative R\&D Environment established to provide access to physics related facilities at the Cavendish Laboratory, University of Cambridge and EPSRC award EP/T00634X/1. SML acknowledges support from EPSRC grant EP/X525686/1. A dataset supporting this work can be found at \url{https://doi.org/10.17863/CAM.104367}. The authors thank Paul Dastoor for helpful discussions.

\clearpage
\appendix

\setcounter{figure}{0}
\renewcommand{\figurename}{Fig.}
\renewcommand{\thefigure}{A\arabic{figure}}

\setcounter{table}{0}
\renewcommand{\tablename}{Table}
\renewcommand{\thetable}{A\arabic{table}}

\section{Test sample images}
\label{app:sample_images}

\begin{figure}[h]%
    
\centering

\subfloat[]{\label{a}\includegraphics[width=.3\linewidth]{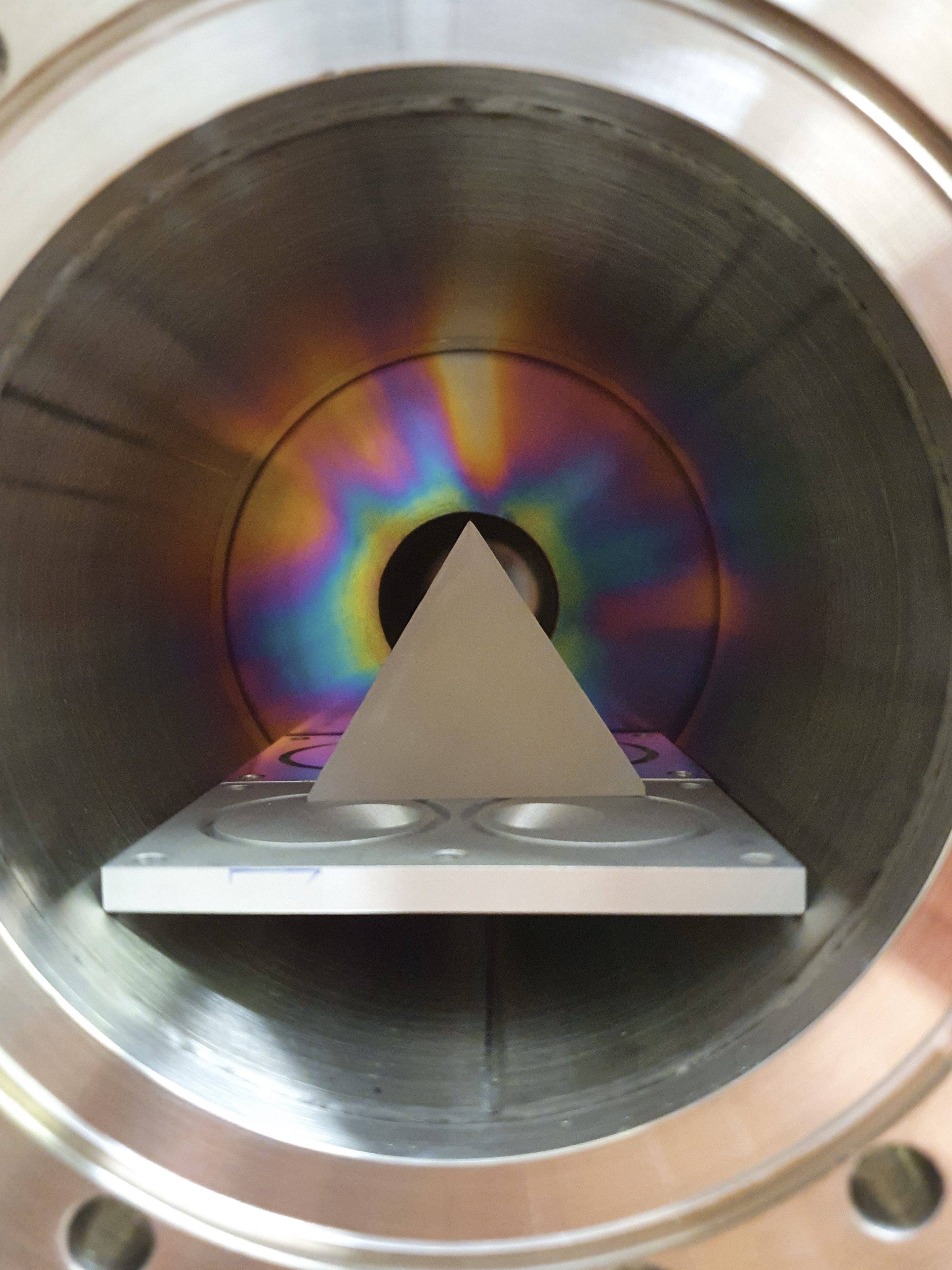}}\par 
\subfloat[]{\label{b}\includegraphics[width=.3\linewidth]{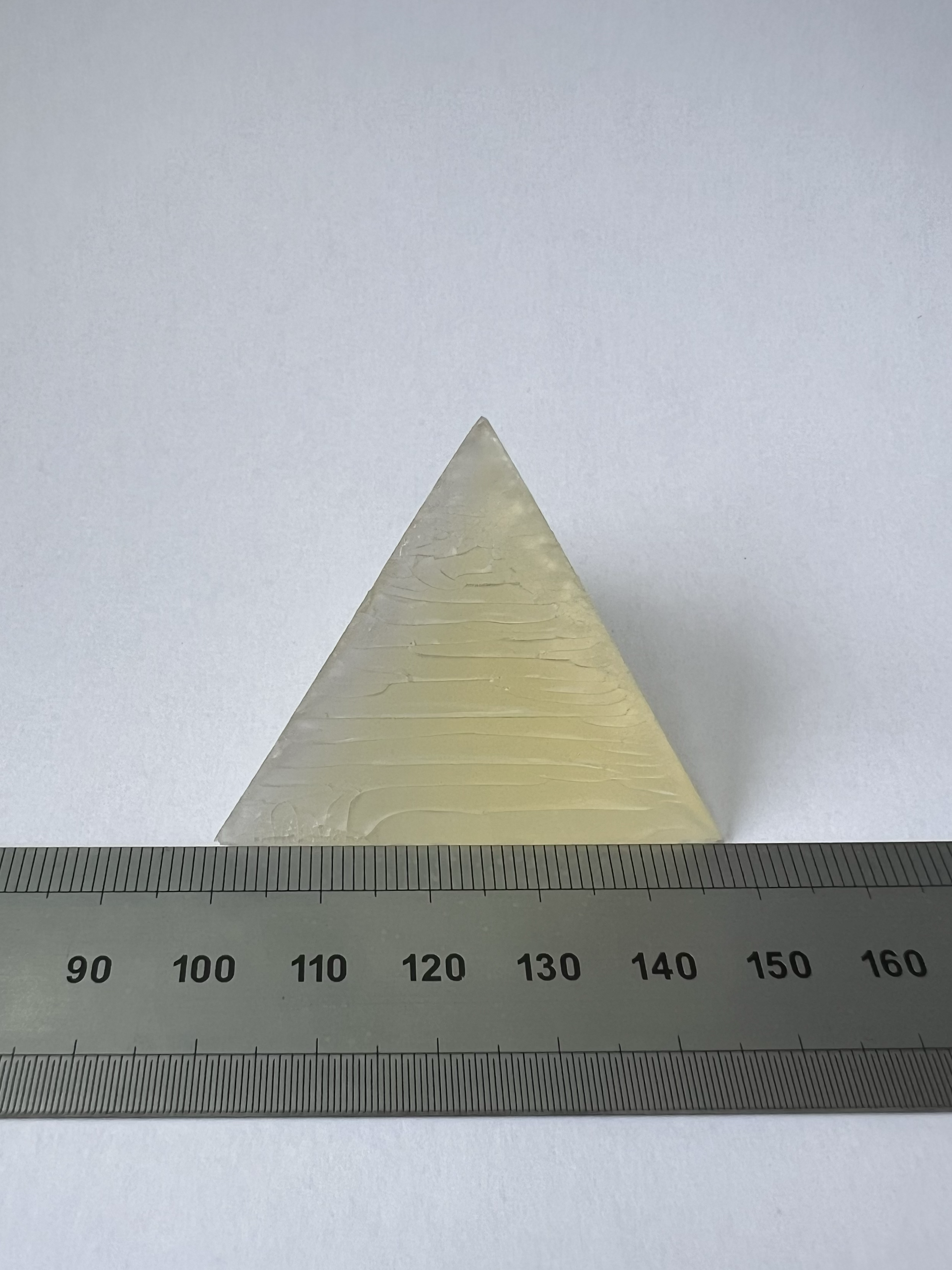}}\qquad
\subfloat[]{\label{c}\includegraphics[width=.3\linewidth]{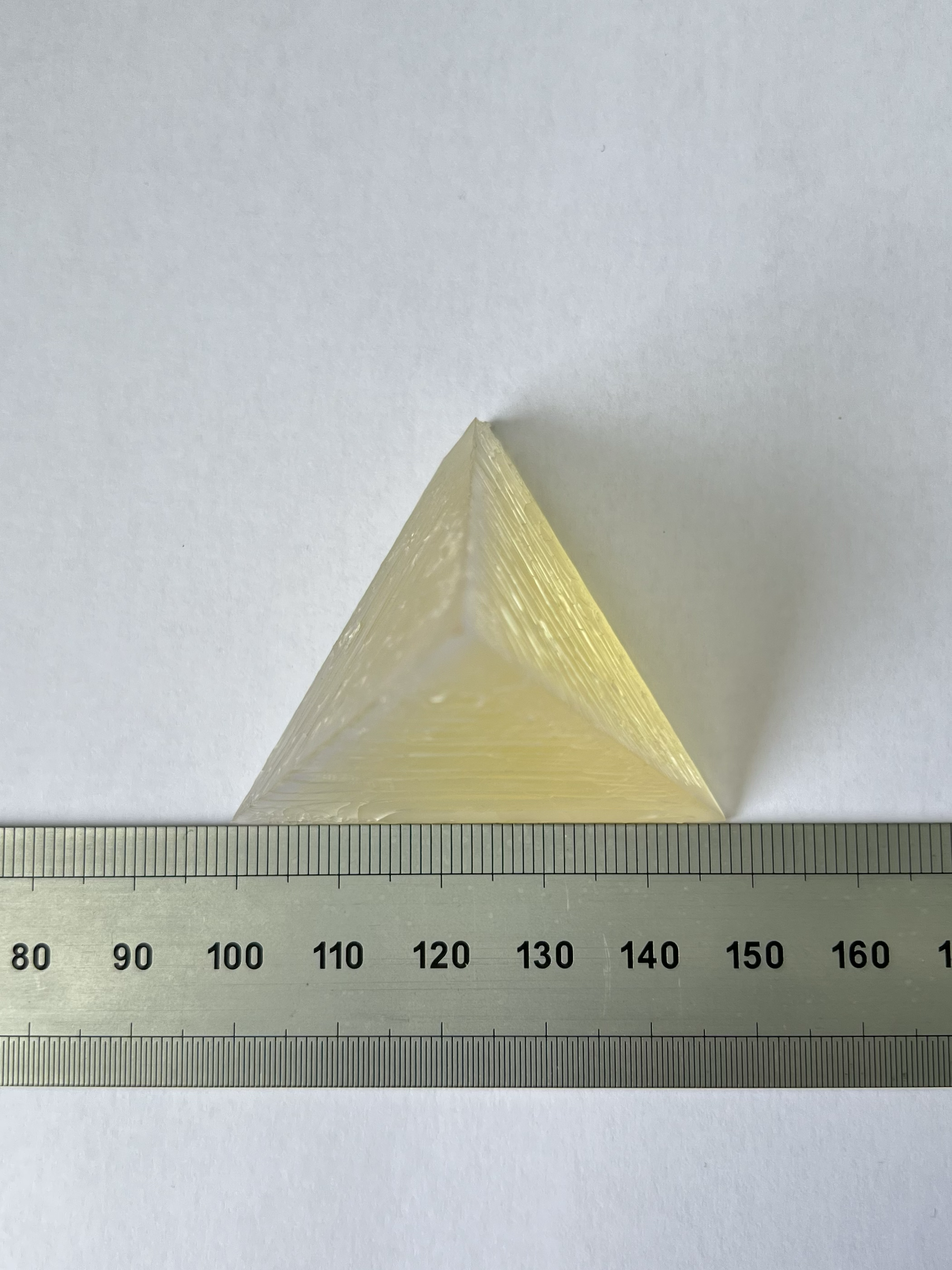}}
\caption{Panel (a), image of test sample pre-baking in the UHV test chamber. Panels (b,c), show the sample that cracked during baking at $\SI{170}{\degreeCelsius}$ with a millimetre scale ruler for reference.}
\label{fig:sample_images}
\end{figure}

\setcounter{figure}{0}
\renewcommand{\figurename}{Fig.}
\renewcommand{\thefigure}{B\arabic{figure}}

\setcounter{table}{0}
\renewcommand{\tablename}{Table}
\renewcommand{\thetable}{B\arabic{table}}

\newpage
\section{High-vacuum test chamber schematic}
\label{app:hv_chamber_schematic}

\begin{figure}[h]
    \centering
    \includegraphics[width=0.5\linewidth]{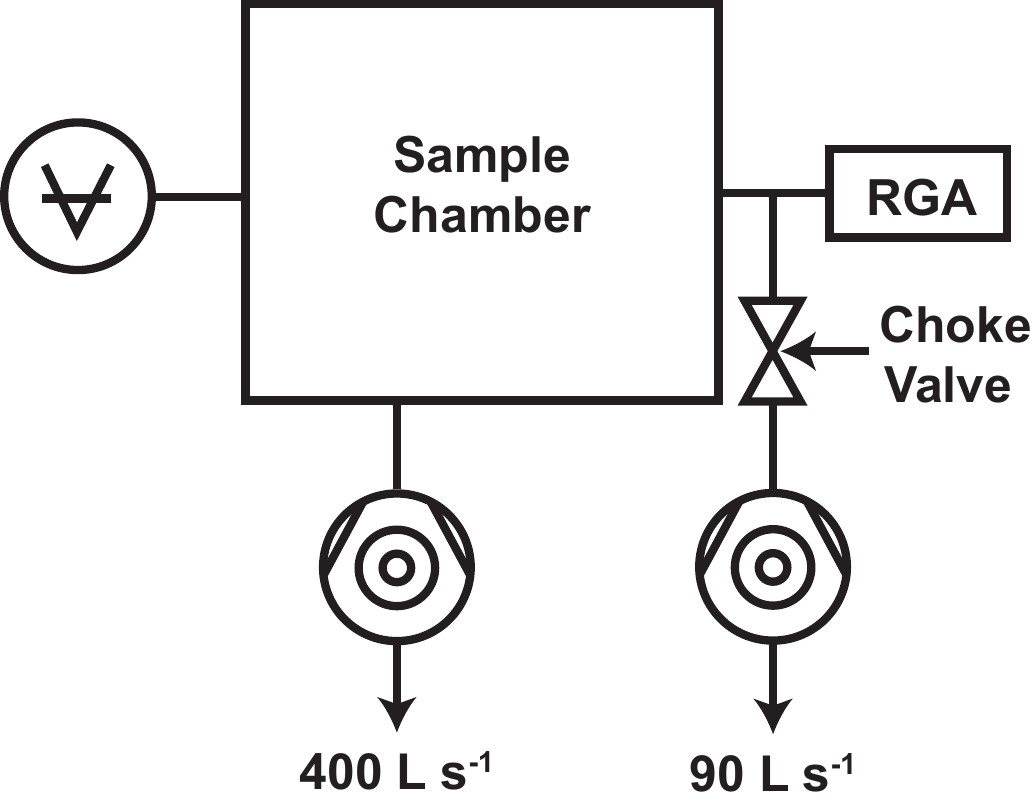}
    \caption{Schematic of the high-vacuum testing system used to collect all data presented. Ion gauge was used to measure pump down curves, from which pressure contributions for each sample were calculated, as shown in Figures \ref{fig:outgassing_plot},\ref{fig:outgassing_abs}. Residual gas analyser (RGA) was used to collect mass spectra in Figures \ref{fig:stacked_rga_tetra}, \ref{fig:stacked_rga_plates}. Total pumping speed out of the sample chamber is approximately $\SI{450}{\litre\per\second}$ due to a choke valve, variable between $2-\SI{50}{\litre\per\second}$, fitted on the turbomolecular pump to restrict its pumping out of the RGA. N\textsubscript{2} pumping speeds quoted. Pressure gauge is an Edwards Active Ion Gauge (AIGX).}
    \label{fig:hv_chamber_schematic}
\end{figure}

\setcounter{figure}{0}
\renewcommand{\figurename}{Fig.}
\renewcommand{\thefigure}{C\arabic{figure}}

\setcounter{table}{0}
\renewcommand{\tablename}{Table}
\renewcommand{\thetable}{C\arabic{table}}

\section{Ultrahigh-vacuum baking/test chamber schematic}
\label{app:uhv_chamber_schematic}

\begin{figure}[h]
    \centering
    \includegraphics[width=0.5\linewidth]{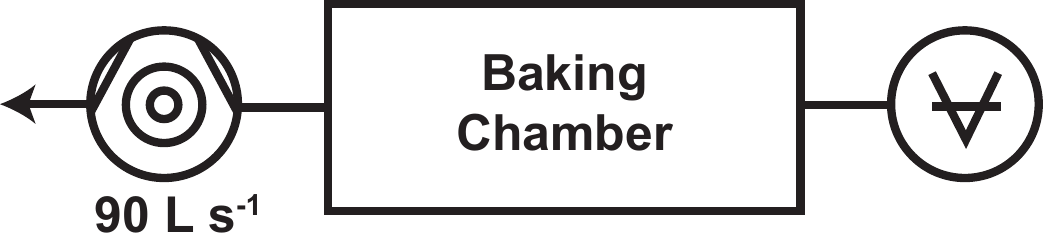}
    \caption{Schematic of the ultrahigh-vacuum test chamber, also used for initial baking of samples. Total pumping speed out of the sample chamber is approximately $\SI{90}{\litre\per\second}$. N\textsubscript{2} pumping speed quoted. Pressure gauge is an Edwards Wide Range Gauge (WRG)\cite{EdwardsGaugeSpecs}.}
    \label{fig:uhv_chamber_schematic}
\end{figure}

\newpage

\setcounter{figure}{0}
\renewcommand{\figurename}{Fig.}
\renewcommand{\thefigure}{D\arabic{figure}}

\setcounter{table}{0}
\renewcommand{\tablename}{Table}
\renewcommand{\thetable}{D\arabic{table}}

\section{Formlabs Clear Resin outgassing rates}
\label{app:outgassing_rates}
\begin{table}[h]
\begin{tabular}{|c|c|c|}
\hline
\begin{tabular}[c]{@{}c@{}}Exposure Time\\ /days\end{tabular} &
  \begin{tabular}[c]{@{}c@{}}Pressure\\  /$\SI{}{\milli\bar}\,(\pm7\%)$\end{tabular} &
  \begin{tabular}[c]{@{}c@{}}Outgassing Rate\\ /$\SI{}{\milli\bar\per\litre\per\second\per\centi\metre\squared}\,(\pm7\%)$\end{tabular} \\ \hline
\rule{0pt}{10pt}0 & $\SI{2.6e-9}{}$ & $\leq\SI{2.9e-8}{}$ \\ \hline
\rule{0pt}{10pt}1 & $\SI{5.1e-8}{}$ & $\leq\SI{5.7e-7}{}$  \\ \hline
\rule{0pt}{10pt}2 & $\SI{7.2e-8}{}$ & $\leq\SI{8.1e-7}{}$  \\ \hline
\rule{0pt}{10pt}4 & $\SI{1.0e-7}{}$ & $\leq\SI{1.1e-6}{}$  \\ \hline
\rule{0pt}{10pt}7 & $\SI{1.1e-7}{}$ & $\leq\SI{1.2e-6}{}$  \\ \hline
\end{tabular}
\caption{Tabulated pressure ($\SI{}{\milli\bar}$) contributions of each sample to total chamber pressure and specific outgassing rate ($\SI{}{\milli\bar\per\litre\per\second\per\centi\metre\squared}$) measured after $\SI{21}{\hour}$ pumping down in the high vacuum test chamber using an Edwards Active Ion Gauge (AIGX), vacuum system schematic shown in Figure \ref{fig:hv_chamber_schematic}. Pressure data presented in Figures \ref{fig:outgassing_plot},\ref{fig:outgassing_abs}. The calculation method for outgassing rates is described in Appendix \ref{app:outgassing_rates}. Uncertainties have been calculated by propagation in quadrature, starting with manufacturers' uncertainty figures on the pressure gauges ($\pm5\%$)\cite{EdwardsGaugeSpecs} and printing technique used ($\pm\SI{0.15}{\milli\meter}$) \cite{clear_data}. The dominant source of uncertainty is from pressure measurements.}
\label{tab:outgassing_rates}
\end{table}

Outgassing rates throughout the manuscript have been calculated as upper-bounds because the test chambers used (figures \ref{fig:hv_chamber_schematic},\ref{fig:uhv_chamber_schematic}) cannot isolate either chamber all sources of pumping. To calculate the upper-bound, we take the pressure contribution by each sample after $\sim\SI{21}{\hour}$ pumping as indicated by the dashed line in figure \ref{fig:outgassing_plot}, also shown directly in figure \ref{fig:outgassing_abs}, where we make the assumption that pressure is constant in time when it is actually still decreasing for all samples. We multiply the pressure by the effective pumping speed on the test chamber (HV $\sim\SI{450}{\litre\per\second}$, UHV $\sim\SI{90}{\litre\per\second}$), and divide by the sample's surface area ($\sim\SI{39.9}{\centi\meter\squared}$) to arrive at a factor for each test chamber that converts a pressure contribution to an upper-bound outgassing rate. For the HV chamber, the factor is $\times\SI{11.3}{\litre\per\second\per\centi\metre\squared}$, and $\times\SI{2.3}{\litre\per\second\per\centi\metre\squared}$ for the UHV chamber.

\begin{comment}
    
\section{Supplementary Data}
Supplementary data to this article can be found online at \url{https://doi.org/10.17863/CAM.104367} upon publication.
\end{comment}
\newpage
\bibliography{ref}

\end{document}